\documentclass[pdflatex,sn-mathphys-num]{sn-jnl}

\usepackage{siunitx}
\usepackage{amsmath}
\newcommand{\sket}[1]{{\ensuremath{\lvert#1\rangle}}}
\newcommand{\lket}[1]{{\ensuremath{\left\lvert#1\right\rangle}}}
\newcommand{\ket}[1]{\if@display\lket{#1}\else\sket{#1}\fi}
\newcommand{\da}{\dagger}
\newcommand{\sbra}[1]{{\ensuremath{\langle#1\rvert}}}
\newcommand{\lbra}[1]{{\ensuremath{\left\langle#1\right\rvert}}}
\newcommand{\bra}[1]{\if@display\lbra{#1}\else\sbra{#1}\fi}
\newcommand{\cre}[1]{a^\da_{#1}}
\DeclareSIUnit\permille{\text{\textperthousand}}
\DeclareSIUnit\bar{bar}

\theoremstyle{thmstyleone}%
\theoremstyle{thmstyletwo}%

\theoremstyle{thmstylethree}%

\begin{document}

\title[Remote Infrared Absorption Spectroscopy with Undetected Photons]{Remote Infrared Absorption Spectroscopy with Undetected Photons}


\author*[1]{\fnm{Simon} \sur{Neves}}\email{simon.neves@femto-st.fr}

\author[1]{\fnm{Fériel} \sur{Armbruster}}

\author[2]{\fnm{Jean-Pierre} \sur{Wolf}}

\affil[1]{Université Marie et Louis Pasteur, CNRS, institut FEMTO-ST, F-25000 Besancon, France}

\affil[2]{Group of Applied Physics, University of Geneva, rue de l'Ecole-de-Médecine 20, 1205 Geneva, Switzerland}


\abstract{We demonstrate a novel method for remote open-path Fourier-transform infrared spectroscopy with undetected photons. Similar to previous quantum spectroscopy works, a mid-infrared spectrum is reconstructed by detecting a near-infrared radiation only, thus bypassing important limitations of infrared detectors. Our study however relies on the co-propagation of the photon-pair and the pump-laser over the same optical path, which allows the probing of the open atmosphere over long distances. By sending the photons over unprecedented distances of up-to 43.4m in the outdoor atmosphere, we were able to detect butane released in the open-path, as well as natural atmospheric methane, thus demonstrating the first use of infrared spectroscopy with undetected photons for atmospheric measurements.}




\maketitle

\section{Introduction}\label{intro}

Atmospheric concentrations of greenhouse gases (GHGs) have risen significantly over the past two centuries. Methane alone has reached $\SI{166}{\percent}$ above its pre-industrial level, and now accounts for $\SI{16}{\percent}$ of the total GHG-driven radiative forcing responsible for global warming \cite{RadiativeForcingNoaa}. Atmospheric volatile organic compounds (VOCs), although not considered major GHGs, also indirectly contribute to radiative forcing through the formation of tropospheric ozone \cite{curci2009modelling}, and represent a direct threat to human health \cite{montero2018volatile}. Precisely quantifying these emissions and attributing them to their anthropogenic and natural sources is therefore essential to informing climate change mitigation and air quality policies \cite{mcdonald2018volatile,guenther2012model,saunois2019global}. Among the various techniques available, infrared (IR) absorption spectroscopy enables the remote and sensitive tracking of gaseous emissions, making it particularly effective for probing environments which are otherwise inaccessible to in-situ instrumentation. Open-path Fourier-transform infrared spectroscopy (OP-FTIR) relies on a compact interferometer to measure the absorbance of an IR beam propagating through the open atmosphere, thereby retrieving the chemical composition of the probed air column thanks to Beer-Lambert's law \cite{FTIRBook}. A thermal source is typically used, generating a broadband IR radiation, thus enabling the simultaneous monitoring of multiple gas species along the probed path \cite{lin2019application,byrne2020monitoring,you2021quantifying}. Yet such systems are typically limited to open-paths of about $\SI{100}{\meter}$, owing to the low coherence and brightness of the source. Other methods rely on Laser radiation to probe long open-paths, including dual-comb absorption spectroscopy \cite{cossel2017open,giorgetta2021open,westberg2023urban,OP-QCL_JianWeiPan2024} and differential infrared absorption Lidar \cite{robinson2011infrared,innocenti2017differential,titchener2022single,siozos2022autonomous}, but these generally display either a narrower spectral coverage, or significantly greater system complexity. More fundamentally, these techniques all suffer from the limitations of mid-infrared (MIR) photodetectors, which are typically noisy, display low efficiency and require cryogenic cooling.\\

Quantum infrared absorption spectroscopy has recently emerged as a promising alternative to these classical methods, leveraging correlated photons-pairs emitted in nonlinear crystals by spontaneous parametric down-conversion (SPDC). It typically involves spectrally non-degenerate photons travelling in a SU(1,1) interferometer \cite{yurke19862}, with the signal photon being emitted in the visible to near-infrared (NIR) spectral range, and the idler being emitted in the MIR region, where GHGs and VOCs display intense absorption lines. Through the sole observation of the signal-photon's interference occurring by induced coherence in nonlinear crystals \cite{zou1991induced}, the idler's mid-infrared spectrum can be retrieved, whilst the idler photon is discarded \cite{kalashnikov2016infrared,paterova2018measurement}. Hence the need for MIR detectors is completely bypassed, as the interferogram is acquired by a high-efficiency silicon-based detector in the visible/NIR spectral range. Later studies harnessed a Michelson nonlinear interferometric configuration, and reconstructed the MIR spectrum by Fourier transform of the signal's interferogram, with no prior knowledge of the signal-idler spectral correlations \cite{lindner2020fourier,mukai2021quantum,lindner2021nonlinear}. This novel method, refered to as quantum FTIR spectroscopy (QFTIR), has since been the subject of major developments, highlighting its potential for future applications to remote atmospheric sensing. These include highly-sensitive MIR spectroscopy with low probing-power \cite{lindner2023high}, ultra-broadband spectroscopy from $\SI{2}{\micro\meter}$ to $\SI{5}{\micro\meter}$ \cite{ultrabroadbandQFTIR}, and the identification of gaseous mixtures released in ambient air by a $\SI{1.7}{\meter}$-long proof-of-principle open-path quantum FTIR spectrometer (OP-QFTIR) \cite{neves2024open}. An analogous method to QFTIR has also enabled the sensitive and remote sensing of methane, by relying on stimulated parametric down-conversion, although probing a significantly narrower spectral region \cite{dong2025methane,dong2025open}. Despite these promising developments, QFTIR still suffers from critical limitations inherent to the nonlinear Michelson interferometer architecture, which impede its applicability to atmospheric measurements. In particular, this layout features two interferometric arms of equal length, one of which serves both as the idler photon's phase-modulation arm and as the gaseous sample interaction arm. In the context of long open-paths measurements, interferometric arms would scale-up to at least hundreds of meters, making the spectrometer particularly complex to set up, sensitive to air turbulence and optics vibrations, and thus impractical for field deployment. In pursuit of deploying practical OP-QFTIR for remote atmospheric monitoring, a more robust nonlinear interferometer configuration is thus needed, that allows the interaction path to be elongated without compromising the spectrometer's performance. \\

Our work addresses this need, by introducing a practical method for OP-QFTIR spectroscopy, and implementing it to probe the open atmosphere over a maximum path-length of $\SI{43.4}{\meter}$. We constructed a novel nonlinear interferometric layout which tackles the main limitations of current QFTIR spectrometers, by decoupling the idler's phase-modulation from the sample interaction path. The phase-modulation is introduced by a low-footprint interferometric module which is aligned prior to probing the sample. The probing beam is then sent over an arbitrary distance in an open-path, enabling the remote sensing of atmospheric gases. Multiple spectroscopic measurements were performed in an outdoor setting, for different open-path lengths and in a single afternoon, without realignment of the interferometric module. In accordance with theoretical predictions, spectral noise did not display any visible dependence in the interaction length, thus demonstrating the system's robustness to air turbulence and vibrations in return optics. Owing to the broad spectral band of our spectrometer, we demonstrated its use for the remote detection of butane release, and for the analysis of the atmosphere's composition. Atmospheric water vapour and methane could be detected simultaneously in natural concentrations, despite their multiple interfering spectral lines, marking a major milestone in QFTIR spectroscopy. More generally, our results open OP-QFTIR spectroscopy up for concrete applications in the remote monitoring of gaseous atmospheric emissions.\\ 

\section{A Nonlinear Interferometer for Remote Sensing}

\begin{figure}[htbp]
    \centering
\includegraphics[width=1\linewidth]{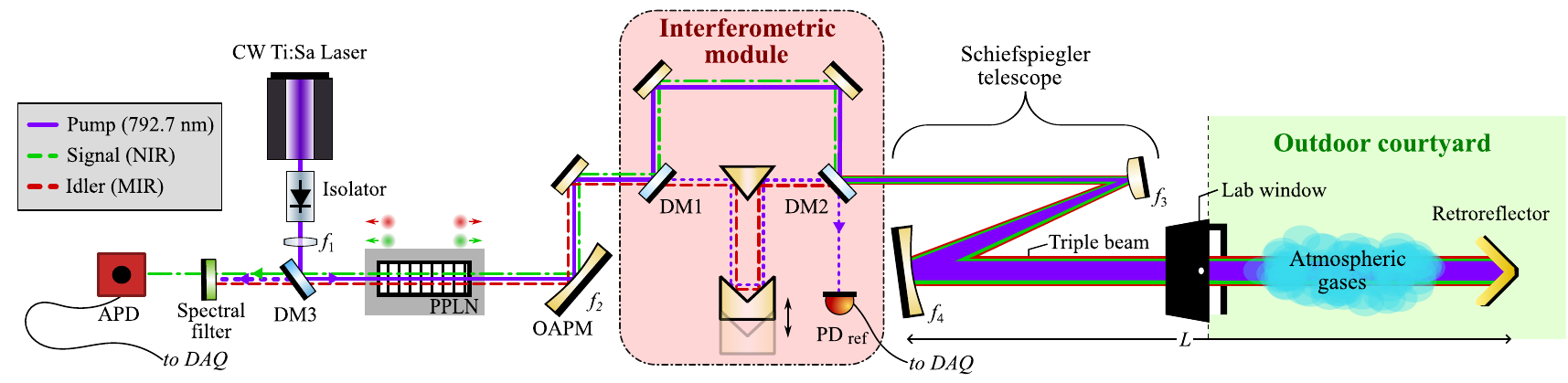}
    \caption{Stable nonlinear interferometer used for remote QFTIR spectroscopy over long open-path. The retroreflector was placed at different locations, at arbitrary distances $L \leq \SI{43.4}{\meter}$ from the telescope's output.}
    \label{fig:Setup Q-FTIR}
\end{figure}

Our study relies on a novel nonlinear interferometric design (see FIG. \ref{fig:Setup Q-FTIR}), which enables measurements over long open-paths. In this layout, a narrowband $(<\SI{1}{\kilo\hertz})$ continous-wave (CW) Titanium-Sapphire (Ti:Sa) laser (Matisse from Spectra-Physics) of $\SI{1.5}{\watt}$ power at $\SI{792.7}{\nano\meter}$ is focused by a $f_1 = \SI{150}{\milli\meter}$ focal-length lens in a nonlinear crystal. The latter is a periodically poled MgO-doped lithium niobate (PPLN) crystal, with a poling period of $\SI{21.8}{\micro\meter}$, a length of $\SI{10}{\milli\meter}$, and stabilized in temperature by an oven. At a temperature of $\SI{75}{\degree\C}$, type-0 quasi-phase matching is achieved for spontaneous down-conversion (SPDC), producing a signal-photon between $\SI{1000}{\nano\meter}$ and $\SI{1050}{\nano\meter}$ in the near-infrared (NIR) region and an idler-photon between $\SI{2650}{\per\centi\meter}$ and $\SI{3050}{\per\centi\meter}$ in the mid-infrared (MIR) region. The measured signal power amounts $\simeq\SI{16}{\nano\watt}$, which corresponds to approximately $\overline{n} = 0.03$ photon pairs emitted per biphoton correlation time $t_c \simeq \SI{0.3}{\pico\second}$, thus in the low gain regime $\overline{n} \ll 1$. In these condition, the biphoton state takes the following form \cite{paterova2018measurement,mukai2021quantum}:
\begin{equation}\label{eq:biphoton_state}
\begin{aligned}
    \ket{\psi} \propto& \iint d\nu_i d\nu_s S(\nu_i,\nu_s)a^\da_{i}(\nu_i)a^\da_{s}(\nu_s)\ket{vac}\\
    = &  \int_0^\infty d\nu_i \Tilde{S}(\nu_i)a^\da_{i}(\nu_i)a^\da_{s}(\nu_p - \nu_i)\ket{vac},
\end{aligned}
\end{equation}
where $\nu_{p,s,i}$ are the respective wavenumbers of the pump, signal and idler, $S(\nu_i,\nu_s)$ is the joint spectral amplitude (JSA) of the biphoton which is determined by the pump Laser's and nonlinear crystal's properties, and $\Tilde{S}(\nu_i)$ is the idler's reduced spectral amplitude, such that:
\begin{equation}
    S(\nu_i,\nu_s) = \delta(\nu_p - \nu_i - \nu_s)\Tilde{S}(\nu_i).
\end{equation}

In this way, the photons' wavenumbers are highly anti-correlated, owing to the pump's narrowband.
At the exit of the nonlinear crystal, the pump, signal and idler are collimated by an off-axis paraboloid mirror (OAPM) of $f_2 =\SI{150}{\milli\meter}$ reflected focal length and sent to the interferometric module. The latter is made of two dichroic mirrors (DM1 and DM2) spliting the idler from the signal and the pump, and recombining the three beams. A tunable retardance is induced on the idler field by a delay-line placed between the two dichroic mirrors, made of two silver-coated prisms with parallel faces, and a $\SI{9.5}{\milli\meter}$ travel-range linear scanning stage. The classical interference of the pump laser's leaks through DM1 and DM2 is measured by a photodiode (PD\textsubscript{ref}), enabling the optimization of the module's alignment, and the precise tracking of the idler's retardance. Remarkably, the interferometric module's footprint amounts to $\SI{20}{\centi\meter}\times\SI{40}{\centi\meter}$ regardless of the open-path's length.\\

\begin{figure}[htbp]
    \centering
\includegraphics[width=0.7\linewidth]{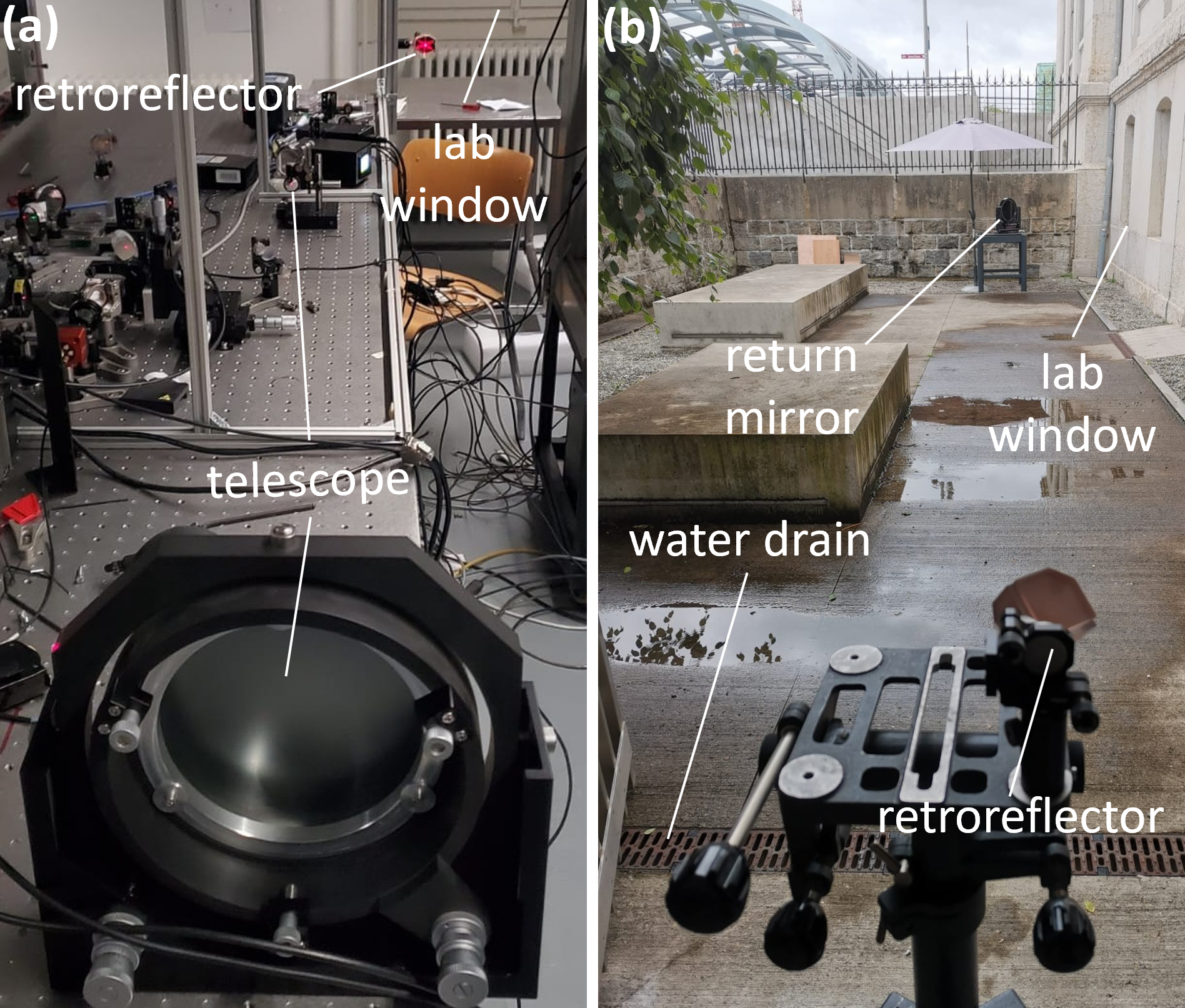}
    \caption{Pictures of (a) the indoor part of the experiment, when the retroreflector is placed inside the lab, (b) the outdoor part of the experiment, when the retroreflector is placed on a tripod outside the lab, \SI{19.8}{m} away from the telescope.}
\label{fig:xpPic}
\end{figure}

Exiting the interferometric module, all three beams follow a common path. The resulting triple beam is expanded by a broadband reflective \textit{Schiefspiegler} telescope \cite{kutter1958schiefspiegler}, made of a $f_3 = -\SI{200}{\milli\meter}$ focal-length convex mirror and a $f_4 = +\SI{1700}{\milli\meter}$ focal-length concave mirror in an oblique configuration. The beam's incidence angles on both mirrors are kept below $\SI{0.1}{\radian}$ to limit optical aberrations. We evaluate the output pump beam's waist around $w_p \simeq \SI{1}{\centi\meter}$, so the beam's divergence is kept below $\SI{0.1}{\milli\radian}$. The triple-beam then interrogate the sample, here the free atmosphere, and is reflected by a silver coated corner-cube retroreflector, placed at an arbitrary distance $L$ from the telescope's output. The reflected light traces back the same path in the opposite direction and passes a second time in the SPDC crystal, where the biphoton's nonlinear interference occurs by induced coherence \cite{zou1991induced}. A $-\SI{60}{\decibel}$ optical isolator protects the pump laser from these returning beams. Another dichroic mirror (DM3) splits the pump from the returning signal photon, which is detected by a NIR avalanche photodiode (APD, A-CUBE-S500-01 from Laser Components). Electronic band-pass filter (SR560 from Stanford Research Systems) with $\SI{1}{\kilo\hertz}$ and $\SI{30}{\kilo\hertz}$ cutoff frequencies, is used to cancel noise from the APD's voltage, which is then recorded by a data acquisition system (DAQ) at a $\SI{100}{\kilo\hertz}$ rate. By moving the linear stage back-and-forth at a $\SI{4}{\milli\meter\per\second}$ speed, we continuously record the signal photon flux $P_s(x)$ for idler's retardance $x \in \ ]-x_0,+x_0[$, where $x_0$ is the maximal absolute retardance applied to the idler. In the ideal case where $x_0 = +\infty$, the inverse Fourier transform of the interferogram provides the idler's spectral intensity, including the specific absorption bands of gases present in the atmosphere:
\begin{equation}\label{eq:spectral_intensity}
\begin{aligned}
       I(\nu_i) =&\ \biggl\vert\frac{1}{2\pi}\int P_s(x)e^{j2\pi\nu x} dx\biggr\vert\\
        \propto&\ |\Tilde{S}(\nu_i)|^2\cdot T(\nu_p - \nu_i)\cdot T(\nu_i),
\end{aligned}
\end{equation}
where $T(\nu)$ is the atmosphere's transmissivity (the reader can refer to Supplement A \cite{supp_mat} for the mathematical derivation of this expression). Note here the term $T(\nu_p - \nu_i)$ emerges from signal-idler spectral correlations, and implies that extra-absorption bands are present in the reconstructed MIR spectrum, due to absorption of the signal photon. Because the atmosphere is mostly transparent in the $\SI{1000}{\nano\meter}$ to $\SI{1050}{\nano\meter}$ region, these are ignored in our study. Finally, we retrieve the absorbance by taking the ratio between two spectral intensities:
\begin{equation}\label{eq:absorb}
       \mathcal{A}(\nu) = -\ln\biggl(\dfrac{I_1(\nu)}{I_0(\nu)}\biggr) = -\ln\biggl(\dfrac{ T_1(\nu)}{T_0(\nu)}\biggr),
\end{equation}
where $I_0$ and $I_1$ are two spectral intensities taken in different conditions, $I_0$ generally considered as the reference. For instance, $I_1$ is recorded at a different time or with a different arm's length, in order to detect variations of gas concentrations over time or along the probing arm.\\

In practice, only a portion of the interferogram is recorded, so $x_0$ takes a finite value. In our case, as the idler photon passes twice in the delay-line, $x_0 = \SI{19}{\milli\meter}$. The effective spectral intensity  $\Tilde{I}(\nu)$ thus includes a window function $w$:
\begin{equation}\label{eq:finiteInterf}
       \Tilde{I}(\nu) =\frac{1}{2\pi}\int_{-x_0}^{+x_0} P_s(x) w(x)e^{j2\pi\nu x}dx = [I \ast H](\nu),
\end{equation}
where $H$ is the inverse Fourier transform of $w$, and $I \ast H$ is the convolution product of $I$ and $H$. This results in a minimum $\SI{0.53}{\per\centi\meter}$ spectral resolution, for $w=1$. 
The measured absorbance thus takes a more complex expression, which can be approximated under the assumption of low absorption $\mathcal{A}(\nu) < 0.5$ \cite{DOASBook}: 
\begin{equation}\label{eq:theo13}
       \Tilde{\mathcal{A}}(\nu) = -\ln\biggl(\dfrac{\Tilde{I}_1(\nu)}{\Tilde{I}_0(\nu)}\biggr) \simeq [\mathcal{A} \ast H](\nu). 
\end{equation}
To demonstrate the accuracy of that method, we first applied it to the absorbance measurement of known reference gases, namely n-butane and methane, placed in a gas cell and with a short optical path of $\approx\SI{10}{\centi\meter}$ after the interferometric module. These results are displayed in Supplement B, and are in good accordance with the expected absorbance deduced from the HITRAN database \cite{gordon2026hitran2024}.\\

This novel layout presents several advantages for long-distance atmospheric measurements compared to a nonlinear Michelson interferometer, used in previous quantum FTIR spectroscopy studies \cite{lindner2020fourier,mukai2021quantum,lindner2021nonlinear,lindner2023high,ultrabroadbandQFTIR,neves2024open}. The key novelty resides in the interferometric modulation being realized first, and both signal and idler photons traveling together through the sample. In particular, the interferometric module's alignment is independent of the retroreflector location, which drastically facilitates setting up arbitrary open-path's lengths. This enables practical measurements of the absolute absorbance of an atmosphere section, as demonstrated in later paragraphs. Comparatively, a nonlinear Michelson interferometer is unsuited for that task, due to the whole realignment required when setting up the interferometer's arms length. Additionally, our novel layout is remarkably stable, due to the compactness of the modulation part, and the intrinsic phase stability of the probing arm. Indeed, as all three beams travel through the same open-path, and owing to their phase relations, the nonlinear interference is barely affected by air turbulence or variations of the path's length, such as those caused by shaking of the retroreflector. This property, analogous to Sagnac interferometer's phase stability \cite{kim2006phase}, is theoretically derived in Supplement A \cite{supp_mat}, and further experimental evidence is provided in the following paragraphs. 

\section{Open-Path Quantum FTIR Spectroscopy}
Our novel interferometer was implemented by placing the retroreflector at different locations, away from the telescope's output. The nearest spot was located $L=\SI{3.0}{m}$ away from the output, inside the GAP-Biophotonics laboratory of Geneva University (see FIG.~\ref{fig:xpPic}.a). All other measurements were carried out by sending the triple-beam by the lab's open window, in the building's courtyard (see FIG.~\ref{fig:xpPic}.b). A return mirror was used to aim the beam at the center of the retroreflector. The latter was mounted on a tripod, and placed at different locations in the courtyard. Photons thus traveled a maximum one-way distance of $L=\SI{43.4}{\meter}$ in the open atmosphere. Remarkably, no realignment of the interferometric module or the telescope was required when elongating the probing arm, so the nonlinear interference could be optimized simply by aiming the triple-beam correctly at the retroreflector's center. This allowed us to perform multiple spectrum acquisitions, for different probing arm's lengths, in a single-afternoon measurement-campaign. This breakthrough was practically out of reach by using a nonlinear Michelson interferometer, as a full realignment was required every time the arms' length was modified.\\

\begin{figure}[htbp]
    \centering
\includegraphics[width=0.76\linewidth]{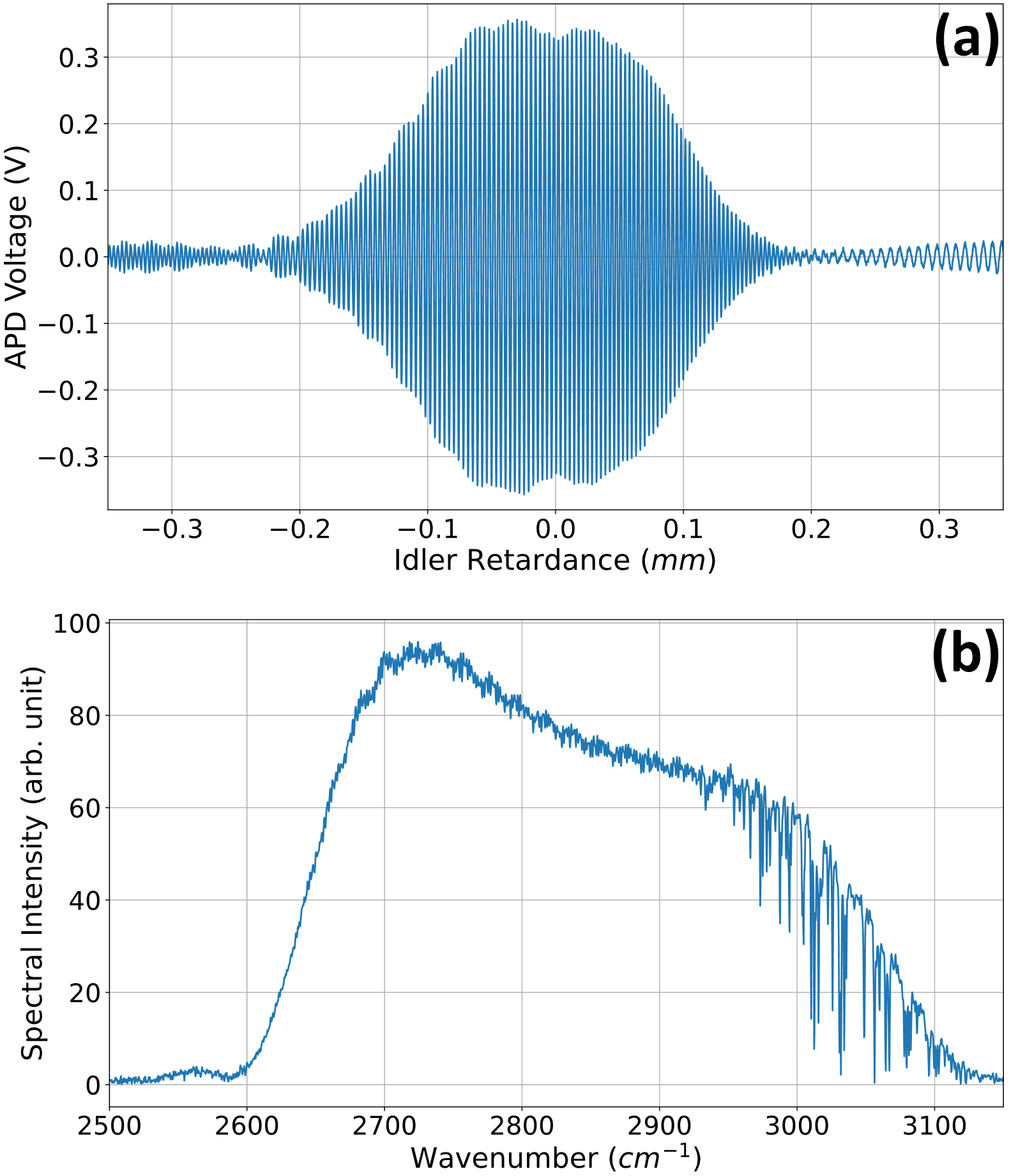}
    \caption{(a) Signal-photon interferogram averaged over 100 iterations, for biphotons traveling over a \SI{43.4}{m} open-path in the atmosphere, and interfering in our novel interferometer. (b) Corresponding Fourier-reconstructed infrared spectrum.}
    \label{fig:recordSpect}
\end{figure}

Each acquisition lasted $\approx\SI{30}{\minute}$, corresponding to 500 successive acquisitions of the interferogram, which are then averaged over $n = 10$ or $n = 100$ iterations in order to increase the SNR of Fourier-reconstructed spectra. Additionally, spectra were apodized using Hann's window \cite{blackman1958measurement,naylor2007apodizing}:
\begin{equation}\label{eq:hann}
           w(x) = \cos^2\biggl(\dfrac{\pi x}{2x_0}\biggr),
\end{equation}
which increases the SNR at the expense of a moderate resolution loss, and limits the overlap of different gases' absorption lines. We display in FIG.~\ref{fig:recordSpect} the interferogram and corresponding infrared spectrum measured for $L=\SI{43.4}{m}$, and averaged over 100 iterations. This result constitutes the record for the longest section of atmosphere interrogated by quantum FTIR spectroscopy. On such a long path, most water vapour absorption lines saturate the spectrum in the $2950-\SI{3100}{\per\centi\meter}$ region. We therefore restrict the rest of the study to shorter open-paths.\\

\begin{figure}[htbp]
  \centering
\includegraphics[width=1\linewidth]{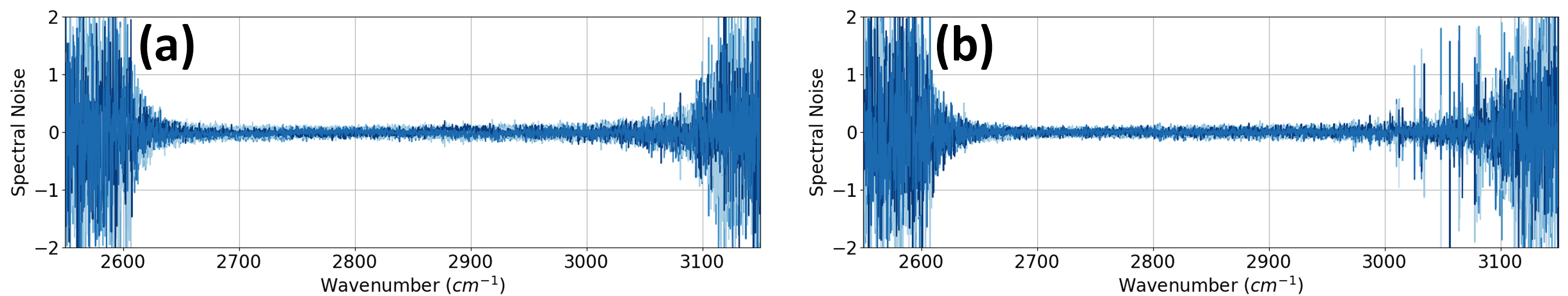}  
\caption{Spectral noise in the Fourier-reconstructed infrared spectrum, averaged over $n=10$ scans, for photons traveling over $L=\SI{3.0}{m}$ (a) and $L=\SI{43.4}{m}$ (b) in the open atmosphere.}
\label{fig:SpectralNoise}
\end{figure}

\begin{figure}[htbp]
    \centering
\includegraphics[width=0.7\linewidth]{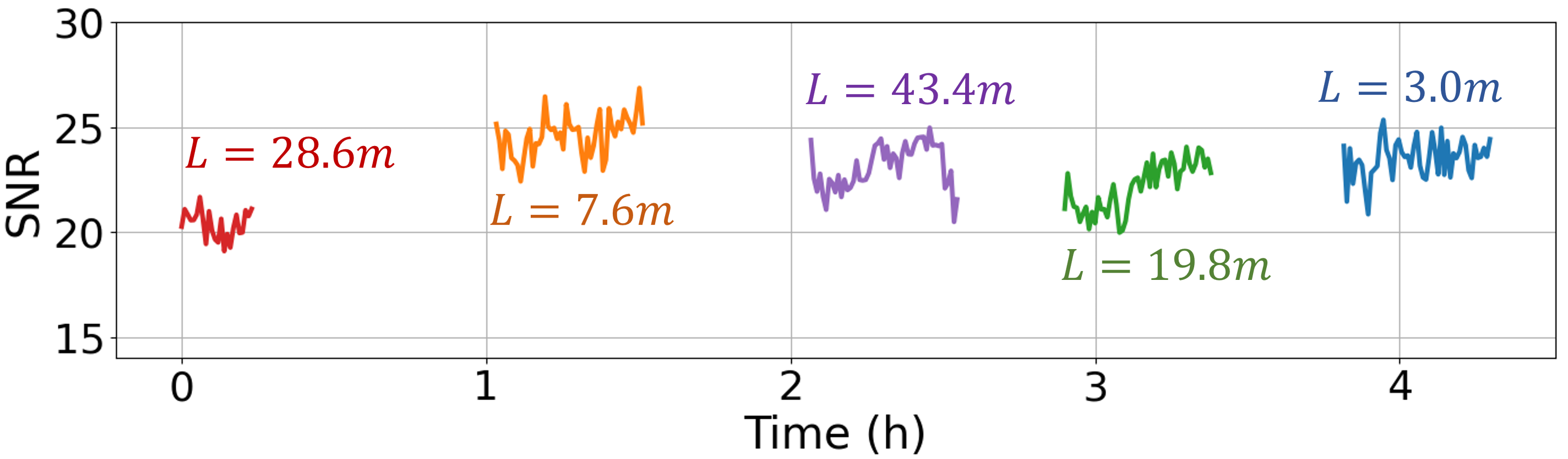}
    \caption{Evolution of the SNR during the 22-05-2025 afternoon measurement campaign. The open-path's length $L$ is specified for each data set. }
    \label{fig:SNRvsTime}
\end{figure}

From the data acquired during the whole measurement campaign, we compute the spectral noise \cite{gattinger2025quantum}:
\begin{equation}
    \mathcal{N}(\nu) = 1 - \Tilde{I}_1(\nu)/\Tilde{I}_0(\nu),
\end{equation} 
where $\Tilde{I}_0$ and $\Tilde{I}_1$ are averaged over 2 subsequently recorded sets of $n$ scans. The corresponding SNR is then given by $SNR = 1/\sigma_\mathcal{N}$, where $\sigma_\mathcal{N}$ is the standard deviation of $\mathcal{N}(\nu)$ over the spectral range of interest. In FIG.~\ref{fig:SpectralNoise} we display the spectral noise for $L=\SI{3.0}{m}$ and $L = \SI{43.4}{m}$, averaging on $n=10$ scans. As expected thanks to the phase-stability of our interferometer, no noise is induced by increasing the open-path's length, except in the $3000-\SI{3050}{cm^{-1}}$ region where narrow water-vapour lines saturate the absorbance. To further investigate our interferometer's stability and the influence of open-path length, we track the average SNR over the $2700-\SI{2900}{cm^{-1}}$ region where the atmosphere is transparent, during the 22nd of May 2025 afternoon measurement campaign (see FIG.~\ref{fig:SNRvsTime}). No realignment of the modulation part or the telescope was performed during this campaign, even when setting up the probing arm's length. The SNR was consistently maintained around a value of $SNR = 23\pm3$, regardless of the open-path's length. This demonstrates the remarkable stability of our setup, even when elongating the open-path, and thus its applicability to the remote sensing of atmospheric gases.

\subsection{Remote Monitoring of a Gas Release}

This novel layout is particularly suited for the remote detection of sudden gas concentration changes at an arbitrary distance from the spectrometer. A typical application of such a measurement is the detection of gas leakage on a pipeline. We simulate such a measurement by remotely sensing n-butane released in the atmosphere by a blowtorch on the triple-beam's path. We place the corner cube in the middle of the courtyard, $\SI{23.3}{\meter}$ away from the telescope, and continuously detect signal photons all-the-while scanning the interferogram 500 times, corresponding to a $\SI{30}{\min}$ acquisition time. The resulting spectral intensity is averaged over $n=10$ scans. The experiment starts at time $t_0$, and no n-butane is released initially. The first 10 scans, acquired from $t_0$ to $t_0 + \SI{36}{\second}$ produce the reference spectral intensity $I_0$, used to compute the measured absorbance $\Tilde{\mathcal{A}}_t(\nu)$ with the spectral intensity $I_t$ taken at time $t$:
\begin{equation}\label{eq:measured}
    \Tilde{\mathcal{A}}_t(\nu)  = -\ln\biggl(\dfrac{I_t(\nu)}{I_0(\nu)}\biggr).
\end{equation}

\begin{figure}[htbp]
    \centering
\includegraphics[width=0.7\linewidth]{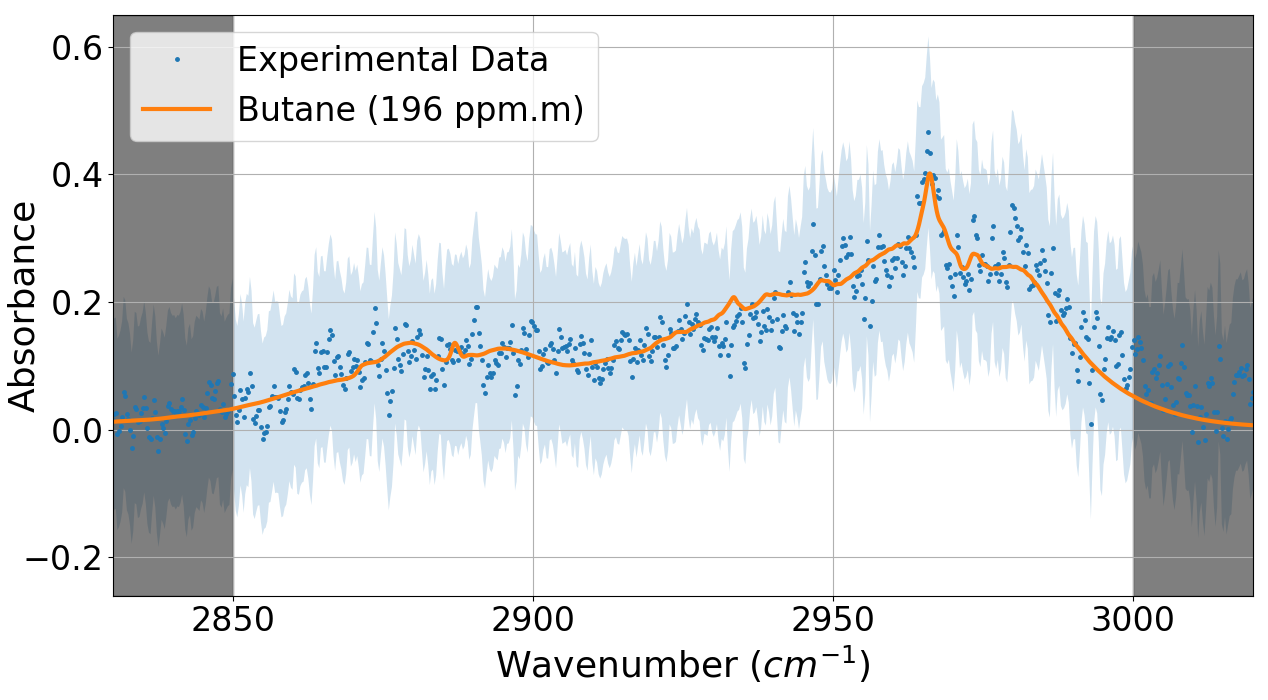}
    \caption{Absorbance spectrum recorded when releasing n-butane in the photons' path, averaged on 10 scans. The absorbance of $\SI{196}{ppm\cdot\m}\pm\SI{74}{ppm\cdot\m}$ n-butane, is obtained by fitting the experimental data, ignoring spectral regions indicated by grey stripes. We provide the absorbance evolution during the experiment in Supplement C \cite{supp_mat}.}
    \label{fig:Butane}
\end{figure}

Acquiring all spectral intensities without changing the open-path's length ensures the absorption of non-saturating absorption lines from atmospheric methane and water-vapour cancel out, so only the n-butane's absorption is measured. At $t_0 + \SI{360}{\second}$, around the 100-th scan, we open the blowtorch, releasing n-butane on the triple-beam's path. We then close the blowtorch around $t_0 + \SI{470}{\second}$. In Fig.~\ref{fig:Butane}, we display the absorbance thus acquired between $t_0 + \SI{414}{\second}$ and $t_0 + \SI{450}{\second}$, averaged on 10 scans. In order to track n-butane's concentration on the triple-beam's path, we decompose the measured absorbance into two terms \cite{DOASBook}:
\begin{equation}\label{eq:BL}
     \mathcal{A}_t(\nu) = \overline{\mathcal{A}}_t(\nu) + L\sum_{k=1}^N \Delta c_k(t)\bigl[\sigma_k*H\bigr](\nu),
\end{equation}
where $\overline{\mathcal{A}}_t(\nu)$ is a baseline typically induced by scattering and variation of photon flux between times $t_0$ and $t$, and the remaining sum is the absorbance of $N$ different gases in the atmosphere as predicted by Beer-Lambert's law, $\{\Delta c_k(t)\}$ are the variations of said gases' concentrations from instant $t_0$, and $\{\sigma_k\}$ are the gases' absorption cross-sections. Note $H$ is again the inverse Fourier transform of the apodization window $w$, emerging from eq.~\ref{eq:theo13}, so eq.~\ref{eq:BL} again assumes a low absorbance $\Tilde{\mathcal{A}} < 0.5$. We focus on the $2850 - \SI{3000}{\centi\meter^{-1}}$ window, where n-butane displays maximum absorbance, and does not overlap with saturating methane or water-vapour lines.\\

Following the same earlier method, we evaluate $SNR = 20$, so a minimum variation of $0.15$ in absorbance can be distinguished from noise with $\SI{99}{\percent}$ confidence. This corresponds to $\SI{74}{ppm\cdot\m}$ path-integrated concentration of n-butane, based on the $\SI{2966}{\centi\meter^{-1}}$-peak. We fit the absorbance using eq.~\ref{eq:BL} as model with n-butane as the single absorbent, and its concentration left as a free parameter. The cross-section is taken from HITRAN database \cite{gordon2026hitran2024}, and we consider a linear baseline. Our data thus shows good accordance with the theoretical absorbance spectrum of n-butane, with $\SI{196}{ppm\cdot\m}\pm\SI{74}{ppm\cdot\m}$ path-integrated concentration. This demonstrate the first remote detection of gas emission in the atmosphere via QFTIR spectroscopy.

\subsection{Practical Detection of Atmospheric Methane}

\begin{figure}[htbp]
    \centering
\includegraphics[width=1\linewidth]{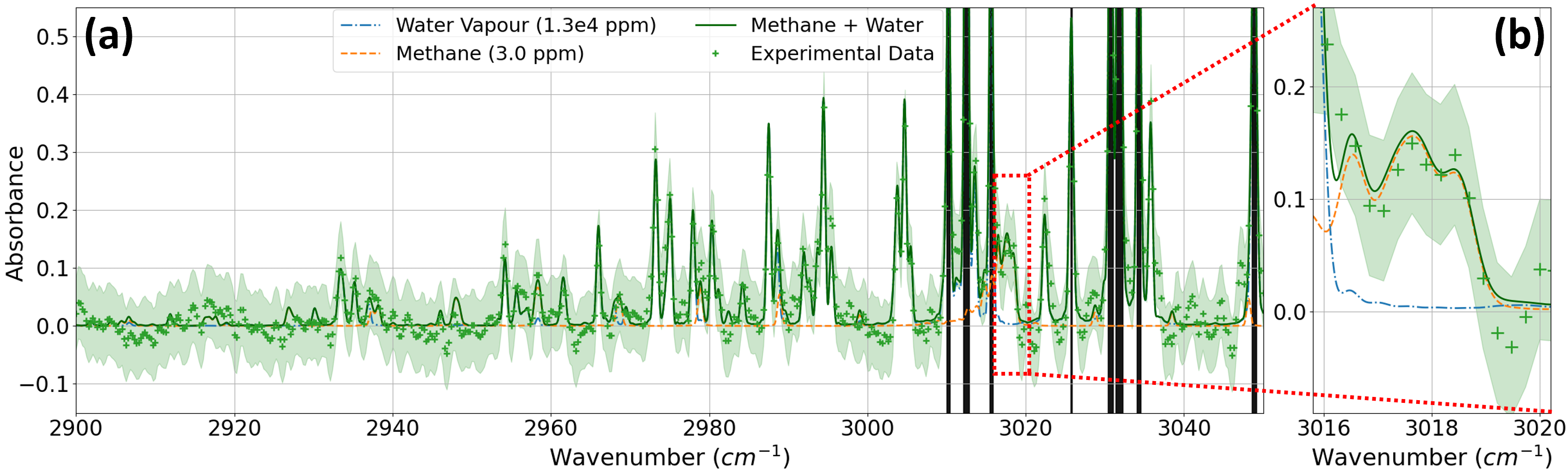}
    \caption{Absorbance spectrum recorded when probing $\SI{21}{\meter}$ of outdoor ambient air. The absorbance of a mixture of $\SI{3.0}{ppm}$ methane and $\SI{1.3e4}{ppm}$ water-vapour is obtained by fitting the experimental data. Grey stripes indicate the ignored data due to water vapour saturation. 
    (a) Full $2900-\SI{3050}{\cm^{-1}}$ region, (b) Zoom on the $3016-\SI{3020}{\cm^{-1}}$ region featuring a specific methane absorption line.}
\label{fig:MethaneAmbient}
\end{figure}

In contrast to the more common nonlinear Michelson interferometer, our novel layout also enables the practical analysis of the atmosphere's composition, owing to the possibility to elongate the probing arm without any interferometric realignment. We demonstrate this property by interrogating a $\SI{21}{m}$-long horizontal section of outdoor atmosphere. We compute the measured absorbance $\Tilde{\mathcal{A}}(\nu)$ (see eq.~\ref{eq:theo13}) between $I_0$ when placing the retro-reflector just outside the lab, $L = \SI{7.6}{\meter}$ away from the telescope, and $I_1$ when placing the retro-reflector $\SI{21}{\m}$ further ($L = \SI{28.6}{\meter}$). Data is averaged over 100 subsequently acquired spectra, and we focus on the $2900-\SI{3050}{\cm^{-1}}$ region, where the most intense atmospheric absorbents are water-vapour and methane. The resulting apodized absorbance is displayed in Fig.~\ref{fig:MethaneAmbient}. We evaluate an average $SNR = 48$ in this spectral region (excluding saturating water-vapour lines), so a minimum variation of 0.063 in absorbance can be distinguished from noise, with $\SI{99}{\percent}$ confidence. This corresponds to variation of $\SI{25}{ppm\cdot\m}$ in methane path-integrated concentration, based on the most intense $\SI{3017.5}{\cm^{-1}}$ absorption line. Assuming methane is evenly spread over the $\SI{21}{m}$ open-path, our spectrometer thus achieves a sensitivity to a minimum of $\SI{1.2}{ppm}$ change in absolute methane concentration. Similarly for water-vapour, we evaluate a sensitivity to a change of $\SI{4.5e4}{ppm\cdot\m}$ path-integrated or $\SI{2.1e3}{ppm}$ absolute concentration, based on the $\SI{2994.5}{\cm^{-1}}$ line. Note that we do not consider the most intense absorption line this time, as multiple water-vapour lines saturate the spectrum.\\

The saturation of the spectrum by water-vapour is an important limitation when attempting to quantify the atmosphere's composition via absorption spectroscopy, as pointed out in \cite{DOASBook}. When $\mathcal{A}(\nu) > 0.5$, most of the approximations made to derive eq.~\ref{eq:BL} are not valid, leading to errors when evaluating concentrations. Thanks to our spectrometer's resolution, we manage to separate methane absorption lines from saturating water-vapour lines (see for instance the $3016-\SI{3020}{\cm^{-1}}$ region in Fig.~\ref{fig:MethaneAmbient}.b). Apodization further limits the interference between saturating lines and the rest of the absorbance spectrum. Hence we simply discard the saturated part of the spectrum (grey stripes on Fig.~\ref{fig:MethaneAmbient}), and again fit the measured absorbance by taking eq.~\ref{eq:BL} as model with water-vapour and methane as absorbents, cross-sections taken from the HITRAN database, a linear baseline, and concentrations as free parameters. Our data thus shows good accordance with the theoretical absorbance spectrum of a mixture of $\SI{1.3e4}{ppm}\pm\SI{0.2e4}{ppm}$ of water-vapour and $\SI{3.0}{ppm}\pm\SI{1.2}{ppm}$ of methane. This methane concentration is higher yet within the same range as the global concentration of $\SI{1.9}{ppm}$ reported for May 2025 \cite{noaaTrendsNOAA}. Note that a slight excess of atmospheric methane is plausible in our experimental conditions, due to the presence on the triple-beam's path of a water drainage system (see FIG.~\ref{fig:xpPic}) connected to the sewers \cite{liu2015methane}. Nevertheless, this result represents a major milestone, as the first detection of methane in ambient air by QFTIR spectroscopy.\\

\section{Conclusion}

We introduced a practical method for remote OP-QFTIR spectroscopy, leveraging a novel nonlinear interferometric architecture adapted to probing the open atmosphere over large distances. The setup is made of a compact interferometric module, which serves to modulate the phase of the MIR idler photon relatively to the pump-Laser and the NIR signal photon, and a triple-beam, made of the pump, signal and idler, which serves as the interaction field. The triple-beam can be aligned and elongated independently of the interferometric module, in order to enhance the sensitivity to gas traces spread over its path. Our results show the remarkable stability of the interferometer, even when the beam is sent over $\SI{43.4}{\meter}$ in the outdoor atmosphere. Owing to the unprecedented interaction length, we were able to demonstrate a sufficient sensitivity to detect natural atmospheric methane, thus providing the first application of QFTIR spectroscopy for analysing the atmosphere's composition. Although further development is required for sub-ppm precision monitoring of background methane, the system can readily be used for the identification of local sources and leaks. In that sense, the remote detection of butane released over the open-path was also demonstrated, showing the potential of QFTIR to monitor gas leaks and pollutant emissions. In view of such applications, future developments may integrate the interferometric module inside a miniaturized system, as was recently done with a nonlinear Michelson interferometer \cite{ramelow2026polymer}. This would enable bringing the OP-QFTIR spectrometer in the field for comparative tests with classical OP-FTIR systems, such as Bruker's OPS \cite{OPS_Bruker}.

\begin{backmatter}
\bmhead{Funding}
This project has received funding from EQUIPEX+ SMARTLIGHT platform (ANR-21-ESRE-0040), and the EIPHI
Graduate School (ANR-17-EURE-0002), and the Région Bourgogne Franche-Comté.

\bmhead{Acknowledgment}
The authors would like to thank A. Djorovi\'c and M. Moret for their help in setting-up the experiments, as well as J. C. Beugnot, M. Chauvet, F. Devaux, and E. Lantz for their support and fruitful discussions.

\bmhead{Disclosures}
The authors declare no conflicts of interest.

\bmhead{Authors contributions}
S.N. and J.P.W. conceptualized the experiment and the methodology. S.N. and J.P.W. performed the experimental investigations with the support of F.A.. S.N. developed the software and analyzed the data, with the help of F.A.. All authors contributed to the interpretation and visualization of the results, as well as writing, reviewing and editing the manuscript. Funding was acquired by J.P.W. and S.N.. 

\bmhead{Data availability}
Data underlying the results presented in this paper are not publicly available at this time but may be obtained from the corresponding author (S. Neves) upon reasonable request.

\bmhead{Supplemental document} See Supplement \cite{supp_mat} for supporting content.
\end{backmatter}


\newpage
\appendix
\begin{center}
\section*{Supplementary Material}
\bigskip

\end{center}
\section{Theoretical Description of the Experiment}

\noindent We hereby derive the mathematical description of our experiment, following similar steps as \cite{paterova2018measurement,mukai2021quantum}. We assume the pump laser to be monochromatic, of single angular frequency $\omega_p$. We also consider the pump is collimated in the crystal, so the Rayleigh length $z_R$ is larger than the crystal's length $z_C$ ($z_R \gg z_C$). In these conditions, we ignore the photons transverse mode. After a first SPDC process, traveling through the same open-path and returning to the crystal for a second SPDC process, the biphoton is in a state of superposition which reads:
\begin{equation}\label{eq:theo1}
\begin{aligned}
    \ket{\psi} =& \ket{\psi_1} + \ket{\psi_2}\\
    =& \alpha\iint d\omega_i d\omega_s S(\omega_i,\omega_s)a^\da_{i1}(\omega_i)a^\da_{s1}(\omega_s)\ket{vac}\\ &+ \alpha e^{j\phi_p}\iint d\omega_i d\omega_s S(\omega_i,\omega_s)a^\da_{i2}(\omega_i)a^\da_{s2}(\omega_s)\ket{vac} + \ket{vac},
\end{aligned}
\end{equation}

where $\ket{\psi_1}$, resp. $\ket{\psi_2}$ is the quantum state if the biphoton is emitted in the first, resp. second SPDC process, $\ket{vac}$ is the vacuum state, $a^\da_{i1(2)}(\omega_i)$, resp. $a^\da_{s1(2)}(\omega_s)$ are idler's, resp. signal's creation operators on the first (or second) SPDC process, $S(\omega_i,\omega_s)$ is the photon pair's JSA, $|\alpha|^2$ is the pump conversion efficiency in the nonlinear crystal, and $\phi_p = \omega_p t_p$ is the pump's phase after propagating and returning to the crystal. Owing to the pump's narrowband, the photons' frequencies are highly anti-correlated, so the JSA takes the following form:
\begin{equation}
    S(\omega_i,\omega_s) = \delta(\omega_p - \omega_i - \omega_s)\Tilde{S}(\omega_i),
\end{equation}
where $\Tilde{S}(\omega_i)$ is the idler spectral amplitude. This way, equation \ref{eq:theo1} simplifies as follows:
\begin{equation}\label{eq:theo3}
\begin{aligned}
    \ket{\psi} =& \alpha\int_0^\infty d\omega_i \Tilde{S}(\omega_i)a^\da_{i1}(\omega_i)a^\da_{s1}(\omega_s)\ket{vac}\\ &+ \alpha e^{j\phi_p}\int_0^\infty d\omega_i  \Tilde{S}(\omega_i)a^\da_{i2}(\omega_i)a^\da_{s2}(\omega_s)\ket{vac} + \ket{vac},
\end{aligned}
\end{equation}
where $\omega_s = \omega_p-\omega_i$ from now on. Note this expression is valid in the low-gain regime, where the conversion efficiency is much lower than unity $|\alpha|^2 \ll 1$. When the interferometer is perfectly aligned, the following relation applies for the creation operators:
\begin{equation}\label{eq:theo4}
   \cre{s2}=e^{-j\omega_s t_s}\Bigl\{[\tau^*(\omega_s)]^2\cre{s1} + \tau^*(\omega_s) r^*(\omega_s)\cre{u1} + r^*(\omega_s)\cre{u2}\Bigr\},
\end{equation}
\begin{equation}\label{eq:theo5}
   \cre{i2}=e^{-j\omega_i t_i}\Bigl\{[\tau^*(\omega_i)]^2\cre{i1} + \tau^*(\omega_i) r^*(\omega_i)\cre{v1} + r^*(\omega_i)\cre{v2}\Bigr\},
\end{equation}
where $t_{i(s)}$ is the propagation time of the idler (signal) photon in the setup, $\tau(\omega)$ is the transmissivity of the sample (here the atmosphere), $\cre{u1(2)}$ and $\cre{v1(2)}$ are creation operators associated with vacuum modes in a beam-splitter model for the sample's absorption \cite{mukai2021quantum}, and $r(\omega) = \sqrt{1-[\tau(\omega)]^2}$ is the corresponding reflectivity. Note that we take into account the signal's absorption as both photons travel through the sample. The average count rate of signal photons is given by $P_s \propto \bra{\psi}E^{(-)}_sE^{(+)}_s\ket{\psi}$, where $E^{(+)}_s(t) \propto \int d\omega e^{-j\omega t}a_{s1}(\omega)$ is the electric field reaching the detector, and $E^{(-)}_s = \bigl(E^{(+)}_s\bigr)^\dagger$. From eqs. \ref{eq:theo3} to \ref{eq:theo5} we thus express the signal photons' detection probability:

\begin{equation}\label{eq:theo6}
       P_s \propto \int_0^\infty d\omega_i|\Tilde{S}(\omega_i)|^2\Bigl[1 + |\tau(\omega_s)|^4 \ + \bigl(\tau^{2}(\omega_s)\tau^{2}(\omega_i)e^{-j(\omega_p t_p - \omega_s t_s - \omega_i t_i)} + c.c.\bigr)\Bigr],
\end{equation}
where $c.c.$ is the complex conjugate of the preceding term in parentheses, and we use the normalization of $\Tilde{S}$, $\int d \omega_i |\Tilde{S}(\omega_i)|^2$ = 1. Now we introduce the distances $L_{s,i,p}$ traveled by the signal, idler, and the pump respectively, during $t_{s,i,p}$. Note that the pump beam follows the same path as the signal photon, so $L_p = L_s = L$. After the modulation part, the idler photon also follow the same path, so $L_i = L + \delta l$, where $\delta l$ is the path difference introduced by the modulation part. We can thus rewrite $P_s$ as a function of $\delta l$:
\begin{equation}\label{eq:theo7}
\begin{aligned}
       P_s(\delta l) \propto A + 2\pi c \int_0^\infty d\nu_i|\Tilde{S}(\nu_i)|^2\bigl(\tau^{2}(\nu_s)\tau^{2}(\nu_i)&e^{-j2\pi \bigl(n(\nu_p) \nu_p - (n(\nu_s) \nu_s + n(\nu_i)\nu_i)\bigr) 2L} \\& \times e^{-j2\pi n(\nu_i)\nu_i 2\delta l} + c.c.\bigr).
\end{aligned}
\end{equation}
where $A = \int d\omega_i |\Tilde{S}(\omega_i)|^2\bigl(1 + |\tau(\omega_s)|^4\bigr)$ is a constant, $\nu_{s,i,p} = \omega_{s,i,p}/(2\pi c)$ is the spectroscopic wavenumber, $n(\nu)$ is the atmosphere's optical index, and $\nu_s = \nu_p - \nu_i$. Note here that lengths $L$ and $\delta l$ are counted twice to account for the round trip of photons. We take $x = 2\delta l$ to simplify the expressions. In addition, under standard pressure and temperature conditions, ambient air is little dispersive, with $n(\nu) \approx 1$. As a first approximation, we can thus rewrite eq.~\ref{eq:theo6}:
\begin{equation}\label{eq:theo8}
       P_s(x) \propto A + 2\pi c \int_0^\infty d\nu_i|\Tilde{S}(\nu_i)|^2\bigl(\tau^{2}(\nu_s)\tau^{2}(\nu_i) e^{-j2\pi \nu_i x} + c.c.\bigr).
\end{equation}
where the first complex exponent in eq.~\ref{eq:theo6} is canceled out due to energy conservation $\nu_p = \nu_s + \nu_i$. We retrieve the known result, that the signal photon's detection-rate varies as the idler's path changes relatively to the pump's and signal's. However, this time the transmissivity of the signal photon $\tau^{2}(\nu_s)$ also appears in the expression, contrary to previous studies where only the idler's transmissivity $\tau^{2}(\nu_i)$ affects the interference. This is due to photons' spectral correlations, and both of them traveling through the same sample. Most importantly, no phase difference is added by variations of the probing arm's length $L$, or by small variations of $n(\nu)$ which tend to cancel out. Hence the interference is only mildly affected by vibrations of the retro-reflector or fluctuations of optical index, though one might expect a bigger sensitivity to such fluctuations as $L$ increases. Our results empirically show empirically that ambient atmosphere's fluctuations do not affect the interference at least up to $\simeq\SI{50}{\meter}$.\\

Finally, recording the whole interferogram  $P_s(x)$ for $x \in ]-\infty;+\infty[$, and computing the inverse Fourier transform allows to retrieve the spectral intensity:
\begin{equation}\label{eq:theo9}
       I(\nu) = \biggl\vert\frac{1}{2\pi}\int P_s(x)e^{j2\pi\nu x} dx\biggr\vert
        \propto |\Tilde{S}(\nu)|^2\cdot T(\nu_p - \nu)\cdot T(\nu),
\end{equation}
where $T = |\tau |^2$ is the transmission of the sample. As mentioned earlier, an additional factor $T(\nu_p - \nu)$ is introduced compared to previous studies. Extra-losses may thus be expected on long distances, due to Rayleigh and Mie scattering \cite{DOASBook}, and absorption lines in the signal's spectral region result in additional absorption lines in the Fourier-reconstructed idler's spectrum. Finally, we measure the absorbance by taking the ratio between two spectral intensities:
\begin{equation}\label{eq:theo10}
       \mathcal{A}(\nu) = -\ln\biggl(\dfrac{I_1(\nu)}{I_0(\nu)}\biggr) = -\ln\biggl(\dfrac{ T_1(\nu_p - \nu)T_1(\nu)}{T_0(\nu_p - \nu) T_0(\nu)}\biggr),
\end{equation}
where $I_0$ and $I_1$ are two spectral intensities taken in different conditions, $I_0$ generally considered as the reference. For instance, $I_1$ is recorded at a different time or with a different arm's length, in order to detect variation of gas quantity over time or over the probing arm.\\

\section{Absorbance of Reference Gases}

\begin{figure}[htbp]
    \centering
\includegraphics[width=1\linewidth]{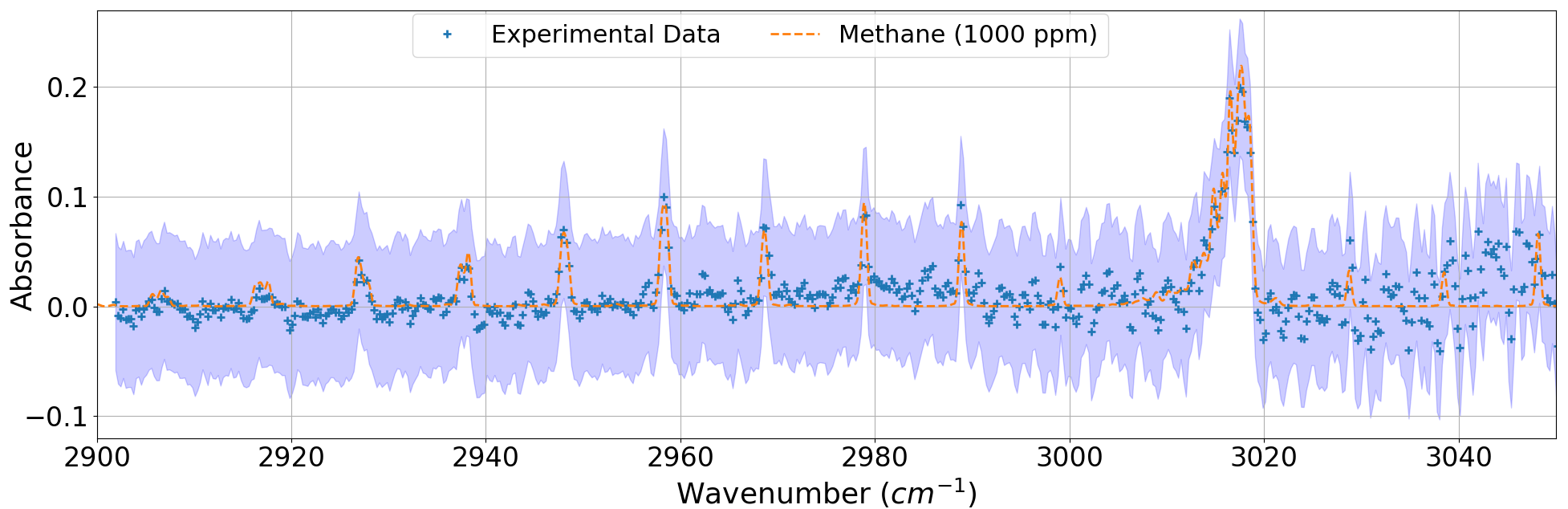}
    \caption{Absorbance spectrum of $\SI{1}{\permille}$ methane in $\SI{1}{\bar}$ of nitrogen, reconstructed with our novel QFTIR spectrometer. The yellow line indicates the expected spectrum taken from the HITRAN-database.}
    \label{fig:methane_ref}
\end{figure}

\begin{figure}[htbp]
    \centering
\includegraphics[width=1\linewidth]{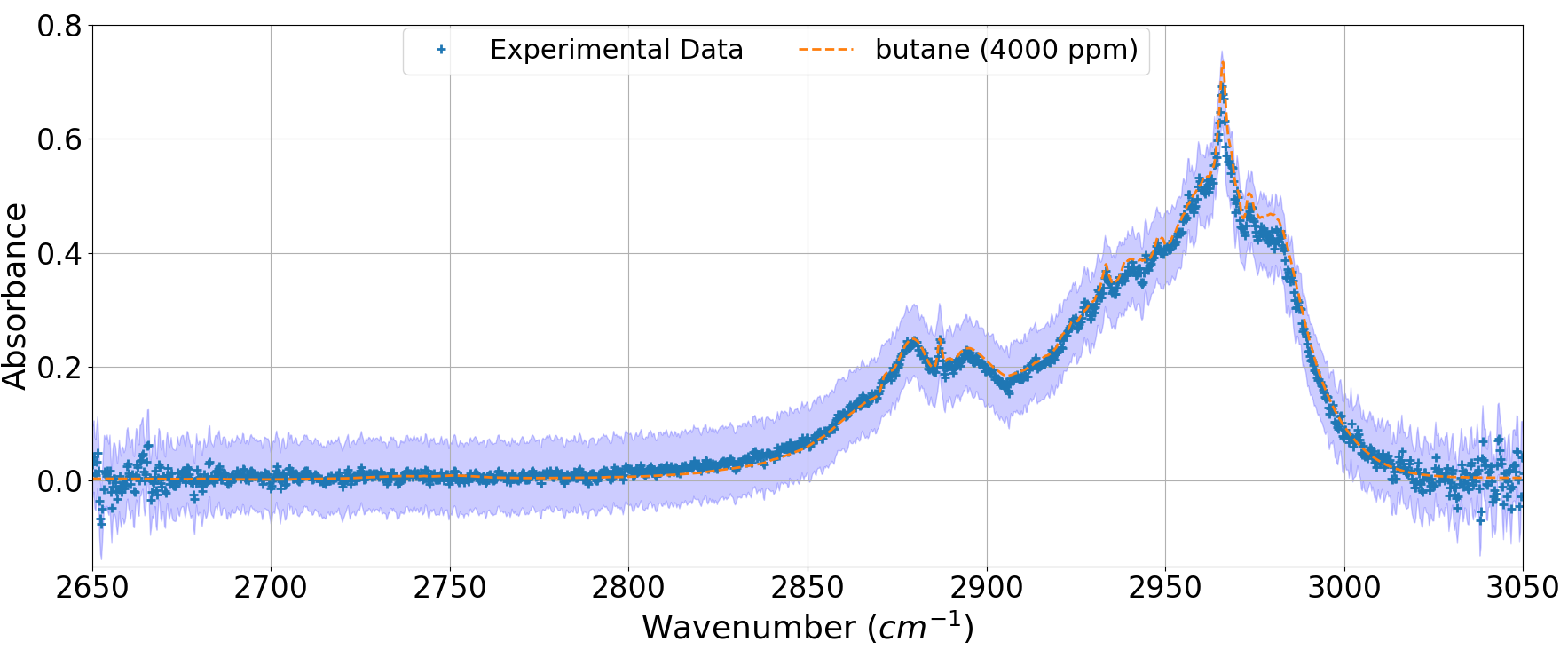}
    \caption{Absorbance of $\SI{4}{\permille}$ n-butane in $\SI{1}{\bar}$ of nitrogen, reconstructed with our novel QFTIR spectrometer. The yellow line indicates the expected spectrum taken from the HITRAN-database.}
    \label{fig:butane_ref}
\end{figure}

In order to demonstrate the accurate measurement of absorbance spectra via our novel QFTIR spectrometer, we measure the absorbance of known reference gases, namely $\SI{1}{\permille}$ methane in $\SI{1}{\bar}$ of nitrogen, and $\SI{4}{\permille}$ n-butane in $\SI{1}{\bar}$ of nitrogen. Each gas is injected inside a $\SI{9}{\centi\meter}$-long gas-cell, placed at the interferometric module's output on the triple-beam's path. The retroreflector is place right after the gas-cell, so the beam travel a short $\approx\SI{20}{\centi\meter}$ path, thus minimizing interference with ambient air's absorbance. Each reference gas' absorbance is obtained by comparing the spectral intensity in these conditions to that obtained when the gas-cell is filled with ambient air. We display in Fig.~\ref{fig:methane_ref} and Fig.~\ref{fig:butane_ref} the absorbance obtained by averaging spectral intensities on 100 consecutive scans. Our experimental data is in good accordance with the expected absorbance spectra taken from the HITRAN-database \cite{gordon2026hitran2024}, which supports our spectrometer's accuracy.  

\newpage

\newpage

\section{Butane Absorbance Evolution}

We hereby provide the absorbance measured continuously while releasing n-butane on the triple-beam's path (see eq. 9 in the main text). Each frame displays the absorbance reconstructed by averaging spectral intensities on 10 consecutive scans, corresponding to a measurement-time of $\SI{36}{\second}$.

\begin{figure}[htbp]
    \centering
\includegraphics[width=0.7\linewidth]{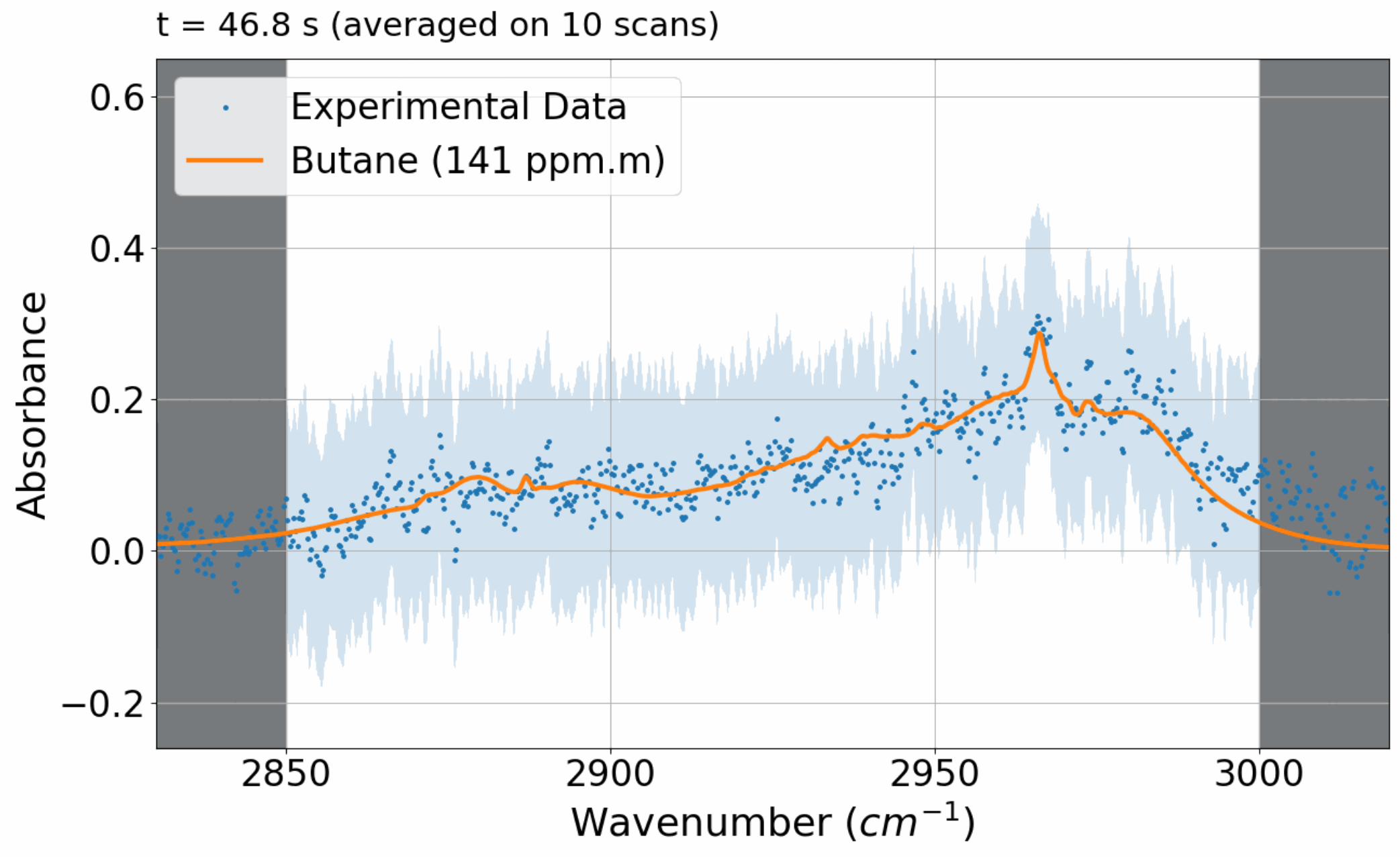}
    \caption{Absorbance evolution while releasing n-butane in the triple-beam's path. Animated absorbance available at \url{https://cloud.femto-st.fr/nextcloud/index.php/s/mL2zPTT8w8gT6jW}}
    \label{fig:butane_evol}
\end{figure}

\end{document}